\documentclass[12pt,a4paper]{article}
\usepackage{times}
\usepackage{epsfig,graphicx}
\usepackage{amsmath}
\usepackage{amssymb}
\usepackage{bm}
\usepackage{enumitem}
\usepackage{xcolor}
\usepackage{booktabs}
\usepackage{tablefootnote}
\usepackage{amsthm}
\newtheorem{proposition}{Proposition}
\usepackage[authoryear,longnamesfirst]{natbib}
\usepackage[a4paper,margin=2.5cm]{geometry}
\usepackage{tabularx}
\usepackage{array}
\usepackage{makecell}

\numberwithin{equation}{section} 
\begin{document}

\title{Parsimonious and Flexible Parameterizations for Realized Covariance Matrix Models}

\author{Edoardo Otranto \\ Department of Social Sciences and Economics, Sapienza University of Rome, \\Piazzale Aldo Moro 5, 00185--Rome, e--mail:edoardo.otranto@uniroma1.it}
\date{}
\maketitle

\begin{abstract}
Covariance models describe the dynamics of conditional covariance matrices as functions of past variances and covariances and are widely used to model the joint evolution of financial asset returns. Their specification faces two major challenges: the curse of dimensionality, whereby the number of parameters increases rapidly with the matrix dimension, and the need to ensure the positive definiteness of the estimated covariance matrices. Moreover, commonly used parsimonious parameterizations, such as scalar and diagonal specifications, fail to capture complex interactions among variances and covariances across assets and cannot accommodate the abrupt changes that often characterize financial markets. We propose a new class of parameterizations that reconciles parsimony with flexibility by allowing distinct, time-varying coefficients for each variance and covariance while requiring only a limited number of parameters. We derive sufficient conditions to ensure the positive definiteness of the resulting covariance matrices and assess the proposed approach through Monte Carlo simulations and an empirical application. The results show that the proposed models can capture complex dependence structures while preserving computational tractability.

\end{abstract}

\section{Introduction}\label{sec:intro}
The availability of detailed financial time-series data, including intraday observations recorded at millisecond frequencies, has fostered the development of high-quality measures of realized variances and covariances \citep{ABCD2013}. These measures are essential for assessing financial market risk and analyzing correlations among assets. Recent approaches have focused on modeling ($n \times n$) realized covariance matrices, including matrices of relatively high dimension.

Two main technical challenges arise when modeling covariance matrices \citep[see, e.g.,][]{BLRsurvey06}: (1) the {\it curse of dimensionality}, since the number of unknown parameters increases rapidly with $n$, given that a covariance matrix has $n(n+1)/2$  distinct elements; and (2) the need to ensure that the model produces positive definite (PD) matrices.

The models proposed in the literature generally extend univariate conditional-variance models, such as GARCH models \citep{Bollerslev_87}, using parameterizations designed to address these two challenges. Examples include scalar and diagonal specifications: in scalar models, all elements of the covariance matrix share the same set of coefficients, whereas in diagonal models, each element follows its own dynamics independently of the others, allowing equation-by-equation estimation. These approaches are commonly adopted in BEKK models \citep{Engle_Kroner95}. However, they offer limited flexibility and do not account for potential interactions among the elements of the covariance matrix.
Another widely used approach is the dynamic conditional correlation (DCC) model of \citet{EngleDCC02}, which involves two estimation stages. In the first stage, conditional variances are estimated using $n$ univariate models; in the second, a scalar GARCH-type model is fitted to the correlation matrix.
  
More recent approaches have been developed specifically for realized measures. One example is the vech-HAR model of \citet{ChiriacVoev_2011}, which extends the HAR model for realized volatility proposed by \citet{Corsi_09},  and has a parametric structure similar to that of the diagonal BEKK model. Another general framework for realized covariance matrices is the conditional autoregressive Wishart (CAW) model proposed by \citet{GGL_2012}. This framework uses the Wishart distribution as the conditional distribution of realized covariance matrices and exploits its properties to obtain consistent estimates. More recent contributions by \citet{Bauwens_Otranto:2023,Bauwens_Otranto:2025} extend the approach of \citet{BauwensOtranto2020} by introducing the Hadamard exponential  parameterization. This parameterization provides greater flexibility by allowing the coefficient associated with each element of the covariance matrix to depend on a small number of unknown parameters, thereby substantially reducing the number of parameters to be estimated.
 An alternative approach, closer to the conventional normal-distribution framework, uses a matrix-variate log-normal distribution \citep[see, e.g.,][]{Gupta_Nagar:2000}. However, the resulting covariance matrix estimator is biased and requires a correction, based either on the ratio of the medians of the actual and estimated standard deviations \citep{Bauer_Vorkink_2011} or on a second-order Taylor expansion \citep{OTRANTO2026}.

In this paper, we develop a general family of models within the CAW framework, based on several alternative parameterizations. The proposed parameterizations ensure the positive definiteness of the estimated covariance matrices. Moreover, the coefficient matrices are time-varying, with distinct coefficients for each element of the covariance matrix, while depending on only a small number of unknown parameters. This parsimonious structure mitigates the curse of dimensionality and makes the models suitable for applications involving relatively high-dimensional matrices. The proposed parameterizations are based on multivariate logistic functions, multivariate linear functions, Hadamard exponential functions, and combinations of these approaches. We impose parameter constraints to prevent explosive dynamics and ensure the positive definiteness of the resulting conditional covariance matrices. We illustrate the approach within the BEKK framework, although it can readily be extended to alternative models by appropriately parameterizing the coefficient matrix associated with the lagged realized covariance matrix. We conduct Monte Carlo experiments to investigate the finite-sample properties of the estimators and their behavior as the time-series length increases. Finally, an empirical application shows that combining the proposed parameterizations can improve model performance.

The remainder of the paper is organized as follows.  Section \ref{sec:modelframe} introduces the CAW framework and the proposed family of models. Section \ref{sec:param}  presents the feasible parameterizations. Section \ref{sec:prop} investigates the properties of the estimators through Monte Carlo simulations. Section \ref{sec:appl} provides an empirical application using the 30 assets included in the Dow Jones Industrial Average (DJIA). Section \ref{sec_concl} concludes. 

\section{The Model Framework}\label{sec:modelframe}
Consider a multivariate time series of $n \times n$  realized covariance matrices, denoted by $\bm{\Gamma}_{t}$, where $t$ is the time index. Let $\mathcal{F}_t= (\bm{\Gamma}_{t}, \bm{\Gamma}_{t-1},\dots)$ denote the information set at time $t$. Conditional on $\mathcal{F}_{t-1}$, $\bm{\Gamma}_{t}$ follows a Wishart distribution with $\nu>n$ degrees of freedom and scale matrix $\bm{S}_t/\nu$, where $\bm{S}_t$ is  symmetric and positive definite. Under this parameterization, $E(\bm{\Gamma}_{t}|\mathcal{F}_{t-1})=\bm{S}_t$. In what follows, we  model the conditional expectation $\bm{S}_t$ using a BEKK representation \citep{Engle_Kroner95} with time-varying coefficients, altough  the proposed approach can readily be extended to other model specifications.

The general model is given by:
\begin{equation}
{\bm S}_t=\bm{K}+\bm{A}_t \odot \bm{\Gamma}_{t-1}+\bm{B}_t \odot \bm{S}_{t-1} \label{eq:tvbekk}
\end{equation}
where $\odot$ is the symbol for the element-by-element  (Hadamard) product, $\bm{K}$, $\bm{A}_t=\left\{a_{ij,t}\right\}$, $\bm{B}_t=\left\{b_{ij,t}\right\}$ are $n \times n$ symmetric matrices of coefficients. The curse of dimensionality is apparent from the fact that these matrices involve $3n(n+1)/2$ different parameters, of which $n(n+1)$ can vary at any time $t$.

The number of coefficients can be reduced by replacing the matrix $\bm{B}_t$ with a scalar $0\le b< 1$; several works show empirical evidence of this result \citep[see, for example,][]{Bauwens_Otranto:2025}.\footnote{We also verified this feature in the application illustrated in Section \ref{sec:appl}.}

A further reduction in the number of unknown coefficients can be achieved through  the covariance targeting specification; this is a simple extension of \citet{Engle_Mezrich96}'s variance targeting idea to the multivariate case \citep[see also][]{noureldin2014}. It consists in reparameterizing the matrix $\bm{K}$ so that the unconditional expectation of ${\bm S}_t$ is equal to the unconditional expectation of $\bm{\Gamma}_{t}$, which we will call $\bar{\bm{\Gamma}}$. For model (\ref{eq:tvbekk}), this could be achieved by replacing $\bm{K}$ with
\begin{equation}
\left[(1-b)\bm{J}_n-\bm{A}_t\right]\odot\bar{\bm{\Gamma}} \label{eq:targ}
\end{equation}
where $\bm{J}_n$ is an $n \times n$ matrix of ones. This substitution has two drawbacks: first, this parameterization does not provide a constant matrix $\bm{K}$; second, and most importantly, (\ref{eq:targ}) is not guaranteed to be PD. One possibility is to adopt the \citet{Hafner:Franses:09} approximation, replacing $\bm{A}_t$  in (\ref{eq:targ}) with a scalar obtained from the average of all its elements at each instant. We propose an alternative parameterization  by introducing a new unknown parameter $\kappa$ and setting 
$\bm{K}=\kappa\bar{\bm{\Gamma}}$, with $\kappa>0$. As in \citet{Hafner:Franses:09}, it does not provide exact covariance targeting because it implies that $E(\bm{S}_t) \ne \bar{\bm{\Gamma}} $, but adds flexibility to the constant part of  the model. The unconditional mean $\bar{\bm{\Gamma}}$ can be consistently estimated in a preliminary step as the sample mean of the realized covariance matrices.

Considering the new parameterization of model (\ref{eq:tvbekk}):
\begin{equation}
{\bm S}_t=\kappa\bar{\bm{\Gamma}}+\bm{A}_t \odot \bm{\Gamma}_{t-1}+b \bm{S}_{t-1} \label{eq:tvbekk_ct}
\end{equation}
the following proposition holds:

\begin{proposition}
  If  $\bm{A}_t$  and the initial value $\bm{S}_0$ are positive semidefinite (PSD),  $\bar{\bm{\Gamma}}$ is estimated by the sample mean  and at least one of the covariance matrices $\bm{\Gamma}_t$ is PD, the matrix $\bm{S}_t$ in (\ref{eq:tvbekk_ct}) is PD. \label{prop1}
\end{proposition}

\begin{proof}
      Obviously, from (\ref{eq:tvbekk_ct}) $\bm{S}_t$ is PSD and, by definition, all the $\bm{\Gamma}_t$ covariance matrices are PSD.  By the Schur product theorem, the Hadamard product of a PD matrix  and a PSD matrix results in a PSD matrix, so $\bm{A}_t \odot \bm{\Gamma}_{t-1}$ is PSD.  If at least one  covariance matrix $\bm{\Gamma}_t$  is PD, then the sample mean $\hat{\bar{\bm{\Gamma}}}$ is PD, and also $\kappa \hat{\bar{\bm{\Gamma}}}$, being $\kappa>0$. It follows that the right-hand side of  (\ref{eq:tvbekk_ct}) is a sum of a PD matrix and two PSD matrices, thus yielding a PD matrix  $\bm{S}_t$, with different coefficients for all the variances and covariances, but with a small set of unknown coefficients.
\end{proof}

The presence of time-varying coefficients implies that (probably) the first and second moments of the process (\ref{eq:tvbekk_ct}) are not constant; however, in the framework of time-varying coefficients, non-stationarity can be modeled \citep[see ][Chap. 17]{lutkepohl:2005}. What we need is to derive some (uniform contraction) condition \citep{Kawano_Hosoe:2024} to rule out explosive dynamics arising from the linear recursion.

\begin{proposition}
A sufficient condition for uniform contraction of the homogeneous
component of the associated time-varying linear recursion is
\[
\sup_t \max_{i,j}|a_{ij,t}+b|<1.
\]
Under the usual regularity conditions, the process is invertible if
$|b|<1$.
\label{prop_si}
\end{proposition}

\begin{proof}
Applying the $\operatorname{vech}$ operator to
(\ref{eq:tvbekk_ct}), which stacks the lower triangular elements of a
symmetric matrix into a vector of dimension $n^*=n(n+1)/2$, we obtain
\begin{equation}
\bm{s}_t=\kappa\bar{\bm{\gamma}}
+\bm{D}_{A_t}\bm{\gamma}_{t-1}
+b\bm{s}_{t-1},
\label{eq:vech}
\end{equation}
where $\bm{s}_t$, $\bar{\bm{\gamma}}$, and $\bm{\gamma}_t$ denote the
vectors obtained by applying the $\operatorname{vech}$ operator to the
corresponding matrices denoted with capital letters, and
$\bm{D}_{A_t}=\operatorname{diag}(\operatorname{vech}(\bm{A}_t))$, where $\operatorname{diag}(\bullet)$ is the operator applied to a vector to create a diagonal matrix with diagonal entries equal to the elements of the same vector. 

Adding and subtracting $\bm{\gamma}_t$ and $b\bm{\gamma}_{t-1}$
on the right-hand side of (\ref{eq:vech}), and using
$\bm{\eta}_t=\bm{\gamma}_t-\bm{s}_t$, yields the following
VARMA-type representation with time-varying autoregressive
coefficients:
\begin{equation}
\bm{\gamma}_t
=\kappa\bar{\bm{\gamma}}
+(\bm{D}_{A_t}+b\bm{I}_{n^*})\bm{\gamma}_{t-1}
-b\bm{\eta}_{t-1}+\bm{\eta}_t,
\label{eq_varma}
\end{equation}
where $\bm{I}_{n^*}$ denotes the identity matrix of dimension $n(n+1)/2$.

Let $\bm{D}_t=\bm{D}_{A_t}+b\bm{I}_{n^*}$. Since $\bm{D}_t$ is diagonal,
its diagonal entries are $a_{ij,t}+b$. Hence, the matrix norm induced
by the vector infinity norm satisfies
\begin{equation}
\|\bm{D}_t\|_\infty
=\max_{i,j}|a_{ij,t}+b| \label{eq:inf_norm}
\end{equation}
Suppose that $\sup_t\|\bm{D}_t\|_\infty\leq\rho<1$. By repeated
substitution, the coefficient matrix multiplying
$\bm{\gamma}_{t-h-1}$ in the associated homogeneous recursion is
$\prod_{k=0}^{h}\bm{D}_{t-k}$. By submultiplicativity of the induced
matrix norm,
\[
\left\|\prod_{k=0}^{h}\bm{D}_{t-k}\right\|_\infty
\leq \rho^{h+1}.
\]
Therefore, the homogeneous component of the associated time-varying
linear recursion decays geometrically, establishing uniform
contraction. By (\ref{eq:inf_norm}), this is verified when $\sup_t \max_{i,j}|a_{ij,t}+b|<1$.

Finally, since the moving-average polynomial in
(\ref{eq_varma}) is $(\bm{I}_{n^*}-b\bm{I}_{n^*}L)$, where $L$ denotes the lag
operator, the innovation part is invertible when $|b|<1$.
\end{proof}

In the next section, we will provide some different feasible parameterizations for the matrix $\bm{A}_t$  that guarantee their PSDness and parsimony. Furthermore, to ensure the positivity of the variances, we impose $0\le b<1$.

\section{Feasible Parameterizations}\label{sec:param}
We propose four parameterizations of the coefficient matrix $\bm{A}_t$ that ensure its  positive semidefiniteness while allowing all its elements to vary over time through a small number of parameters. The proposed specifications are based on logistic, linear, and Hadamard exponential (HE) functions. positive semidefiniteness of $\bm{A}_t$ implies that its diagonal elements are nonnegative,  the resulting conditional covariance matrix  $\bm{S}_t$ has nonnegative diagonal elements.

\subsection{Logistic Parameterization}\label{sec:st_par}
In (\ref{eq:tvbekk_ct}),  $\bm{A}_t$ is reparameterized as follows:
\begin{equation}
\bm{A}_t=a_0+a_1G_{t}
\label{eq:parlog}
\end{equation}
where $a_0>0$, $a_1 \ge 0$, and $(a_0+a_1+b)<1$. $G_{t}$ is a multivariate logistic function, defined as:
\begin{equation}
G_{t}=(\bm{I}_n+\exp(-\phi(\bm{\Gamma}_{t-1}-c)))^{-1}  \label{eq:logistic}
\end{equation}
with $\phi>0$ and $c$ a scalar coefficient, $\bm{I}_n$ denotes the $n$-dimensional identity matrix.

Logistic functions have been used to obtain time-varying coefficients by \citet{Teräsvirta1994}, introducing smooth-transition autoregressive models in a univariate context. The coefficient $c$ acts as a threshold value, while $\phi$ influences the shape of the logistic function, which can vary from a constant straight line to an S-shape to a step function. Potential identification problems arise because, as $\phi$ increases, and after a certain point, the shape of the function does not change (logistic function saturation). We discuss this issue in greater detail in Section \ref{sec:prop}.

The parameterization (\ref{eq:parlog}) ensures the positive definiteness of $\bm{S}_t$ in (\ref{eq:tvbekk_ct}) and, under the constraints on the coefficients, the stability and invertibility of the process.

\begin{proposition}
     In (\ref{eq:parlog})    $\bf{A}_t$ is PSD. \label{prop2}
\end{proposition}

\begin{proof}
    	 Let ${\bm 0}_n$ denote   the $n \times n$ matrix of zeros. Considering the spectral decomposition of  $\bm{\Gamma}_{t}$:
	 \[
	  \bm{\bm{\Gamma}}_{t}=\bm{V}_t\bm{\Lambda}\bm{V}_t'
	  \]
	  with $\bm{\Lambda}$ a diagonal matrix containing the $n$ eigenvalues $\lambda_i \ge 0$ and $\bm{V}_t$ the corresponding matrix of eigenvectors, for the spectral definition of matrix functions \citep[see, for example,][]{Higham:2008} we obtain the spectral decomposition of $\bm{G}_{t}$:
      \[
       \bm{G}_{t}=\bm{V}_t\bm{\Lambda^*}\bm{V}_t'
      \]
      with $\bm{\Lambda}^*$ a diagonal matrix with elements $0<(1+\exp(-\phi(\lambda_i-c)))^{-1}<1$, so that $\bm{0}_n\prec \bm{G}_{X,t} \prec \bm{I}_n$, where the symbol $\prec$ denotes  the Loewner order,  so that $X\prec Y$ if and only if \(Y-X\) is positive definite.\footnote{Similarly, $X\preceq Y$ indicates that $Y-X$ is positive semidefinite.}  This implies the positive definiteness of $\bm{G}_{t}$ and, as a consequence,   $\bm{0}_n\prec \bm{A}_{t} \preceq a_0+a_1\bm{I}_n$ . 
\end{proof}
      
\begin{proposition}

The condition $(a_0+a_1+b)<1$ is sufficient for the uniform contraction of the homogeneous component of the associated time-varying linear recursion and for invertibility of the innovation filter.  \label{prop3}
\end{proposition}

\begin{proof}

Let $a_0+g_{ij,t}$  be the $(i,j)$ entry of  $\bm{A}_t$ in (\ref{eq:parlog}). From Proposition \ref{prop_si}, the eigenvalues of the matrix $\bm{D}_t$ are $(a_0+g_{ij,t}+b)$.

 Proposition \ref{prop2} establishes that $\bm{0}_n\prec \bm{A}_{t} \preceq a_0+a_1\bm{I}_n$; this implies that $|g_{ij,t}|\le a_1$ for each ($i,j$). Thus, uniform contraction is verified if every element on the diagonal of $\bm{D}_t$ is less than, in magnitude, $(a_0+a_1+b)<1$ for every $t$. Due to the non-negativity constraints on the coefficients $a_0$, $a_1$, $b$,  the previous result also implies $0 \le b<1$, which is the invertibility condition derived in Proposition \ref{prop_si}.

\end{proof}

We call model (\ref{eq:tvbekk_ct}) with parameterization (\ref{eq:parlog}) the Log-BEKK model.

\subsection{Linear Parameterization}\label{sec:linpar}
An alternative (simpler) specification can be obtained by parameterizing   matrix $\bm{A}_t$ in (\ref{eq:tvbekk_ct})  as:
\begin{equation}
    \bm{A}_t=\bm{C}-\alpha_a \bm{\Gamma}_{t-1}\\
    \label{eq:att_term}
\end{equation}
 where  $\bm{C}$ is a PD matrix with all diagonal elements (variance coefficients) equal to $\alpha_v$ and all off-diagonal elements (covariance coefficients) equal to $\alpha_c$  ($\alpha_v>\alpha_c\ge 0$; $\alpha_a\ge 0$). 

This parameterization introduces a dependence of $\bm{S}_t$ on the elementwise product of the lagged realized covariance matrix and a coefficient matrix that varies with $\bm{\Gamma}_{t-1}$.

The coefficient $\alpha_a$ captures a negative effect on $\bm{S}_t$ and is consistent with the idea of an attenuation effect proposed by \citet{BOLLERSLEV_Patton_Quaedvlieg:2016} in the univariate case, and subsequently extended to the multivariate case by \citet{BOLLERSLEV_Patton_Quaedvlieg:2018}.\footnote{They use a vech-HAR specification as the base model.} In these works, the authors parameterize the lagged realized variance coefficients as a linear function of the lagged realized quarticity. \citet{CIPOLLINI_Gallo_Otranto2021} demonstrates that realized variance is a valid substitute for quarticity, with the advantage of being more readily available. Therefore, the parameterization (\ref{eq:att_term}) can be considered an extension to the multivariate case of the univariate parameterization proposed by \citet{CIPOLLINI_Gallo_Otranto2021},
reducing the effect of possible measurement errors and providing a time-varying matrix of coefficients $\bm{A}_t$, all different from each other.

The following propositions establish conditions under which $\bm{A}_t$ is positive semidefinite.

\begin{proposition}
     $\bm{A}_t$ is PSD if and only if, for every $\bm{x} \in \mathbb{R}^n$ and for every $t$
\[\alpha_a \bm{x}'\bm{\Gamma}_t \bm{x} \le (\alpha_v-\alpha_c)||\bm{x}||^2+ \alpha_c (\bm{j}_n' \bm{x})^2,
\]
where $||\cdot||^2$ represents the Euclidean norm and $\bm{j}_n$ an $n-$dimensional vector of ones. \label{prop4}
\end{proposition}

\begin{proof}
    From (\ref{eq:att_term}), $\bm{A}_t$ is PSD if and only if $ \bm{x}'\bm{C} \bm{x} \ge \alpha_a \bm{x}'\bm{\Gamma}_t \bm{x}$ for every $\bm{x} \in \mathbb{R}^n$. Rewriting $ \bm{C}$ as:
\begin{equation}
    \bm{C}=(\alpha_v-\alpha_c) \bm{I}_n+\alpha_c \bm{J}_n \label{eq:rewr}
\end{equation}
where $\bm{J}_n=\bm{j}_n \bm{j}_n'$, the left-hand side of the previous inequality is:
\[
\bm{x}'\bm{C} \bm{x}=(\alpha_v-\alpha_c) \bm{x}' \bm{I}_n \bm{x}+\alpha_c \bm{x}'\bm{J}_n\bm{x}=(\alpha_v-\alpha_c)||\bm{x}||^2+ \alpha_c (\bm{j}_n' \bm{x})^2
\]
The last term is derived from
\[
\bm{J}_n\bm{x}=\bm{j}_n(\bm{j}_n' \bm{x})
\]
so 
\[
\bm{x}'\bm{J}_n\bm{x}=\bm{x}'\bm{j}_n (\bm{j}_n' \bm{x})=(\bm{j}_n' \bm{x})^2 
\]
Therefore, the necessary and sufficient condition to ensure positive semidefiniteness  of $\bm{A}_t$ can be rewritten as:
\begin{equation}
\alpha_a \bm{x}'\bm{\Gamma}_t \bm{x} \le (\alpha_v-\alpha_c)||\bm{x}||^2+ \alpha_c (\bm{j}_n' \bm{x})^2. \label{eq_psd_nsc}
\end{equation}
\end{proof}

\begin{proposition}
    A sufficient condition to ensure the positive semidefiniteness of $\bm{A}_t$ is that $\alpha_a \le (\alpha_v-\alpha_c)/\max(\bm{\xi})$, where $\bm{\xi}$ is the vector containing the maximum eigenvalue of each realized covariance matrix $\bm{\Gamma}_{t}$. \label{prop5}
\end{proposition}

\begin{proof}
From (\ref{eq:rewr}) we immediately  note that the largest eigenvalue of $\bm{C}$ is equal to $\alpha_v+(n-1)\alpha_c$ and the remaining $n-1$ eigenvalues are all equal to $\alpha_v-\alpha_c$. $\bm{A}_t$ is PSD if $\bm{C} \succeq \alpha_a \bm{\Gamma}_{t-1}$, that is, if the eigenvalues of this difference are non negative for every $t$. A sufficient condition is that $\alpha_a \max(\bm{\xi})\le (\alpha_v-\alpha_c)$, yielding the constraint for $\alpha_a$ expressed in Proposition \ref{prop5}.
\end{proof}

\begin{proposition}
 Let ${\bm \Gamma}_t$ ($t=1,\dots,T$) denote the observed sample. Under the conditions of Proposition \ref{prop5}, the coefficient matrix $\bm{A}_t$, with parameterization  (\ref{eq:att_term}), is PSD for every $t=1,\ldots,T$. Moreover, if $(\alpha_v+b)<1$, the homogeneous component of the associated time-varying linear recursion is contractive over the observed sample. \label{prop6}
\end{proposition}

\begin{proof}
 Denoting $\gamma_{rs,t}$ as the generic element of the covariance matrix $\bm{\Gamma}_t$, the parameterization (\ref{eq:att_term}) implies that every diagonal element of $\bm{D}_t$ in Proposition \ref{prop_si} is given by $(\alpha_v-\alpha_a \gamma_{rr,t}+b)$ (when referring to the diagonal elements of $\bm{A_t}$) and  $(\alpha_c-\alpha_a \gamma_{rs,t}+b)$ (when referring to the off-diagonal elements of $\bm{A}_t$); $r,s=1,\dots,n$; $r\ne s$. Considering that $\alpha_a \gamma_{rr,t} \ge 0$
\[
    \alpha_v-\alpha_a \gamma_{rr,t}\le\alpha_v
\]
For the off-diagonal elements $|\gamma_{rs,t}|<\max(\bm{\xi})$, and
\[
|  \alpha_c-\alpha_a \gamma_{rs,t}| \le \alpha_c+\alpha_a |\gamma_{rs,t}|\le \alpha_c+(\alpha_v-\alpha_c)=\alpha_v
\]
The condition derived in Proposition \ref{prop_si}  is therefore satisfied by $(\alpha_v+b)<1$.
\end{proof}

The parameter constraints for the linear specification are computed using the maximum eigenvalue observed in the estimation sample. They therefore ensure positive semidefiniteness of the coefficient matrices over that sample, but do not by themselves guarantee the same property for future observations.

We call model (\ref{eq:tvbekk_ct}) with parameterization (\ref{eq:att_term}) the Lin-BEKK model.

\subsection{Hadamard Exponential Parameterization}
\citet{Bauwens_Otranto:2023} introduce the following parameterization of $\bm{A}_t$ for modeling realized covariance matrices based on the HE operator:
\begin{equation}
    \bm{A}_t=\bm{W}_t=a \exp^\odot(\varphi(\bm{P}_{t-1}-\bm{J}_n))  \label{eq:hepar}
\end{equation}
where $\exp^\odot$ represents the HE operator (element-by-element exponential), $\bm{P}_{t}$ is a correlation matrix. The coefficient $a$ is strictly positive, and $\varphi$ is nonnegative ($\varphi=0$ provides a constant coefficient $a$ for all variances and covariances).

The parameterization (\ref{eq:hepar}) implies that the diagonal elements of $\bm{A}_t$ are equal, while the off-diagonal elements are different. In practice, the variance coefficient is constant and equal for all, while the covariance coefficients are specific and time-varying, increasing when there is a greater correlation, at the previous time, between the corresponding assets.
\citet{BauwensOtranto2020} prove that  $\bm{W}_t$ is PD. Furthermore, considering that all the elements of $\bm{W}_t$ are nonnegative and that the highest value is $a$, a sufficient condition for uniform contraction of the homogeneous component of the associated time-varying linear recursion is $(a+b)<1$.

We call model (\ref{eq:tvbekk_ct}) with parameterization (\ref{eq:hepar}) the HE-BEKK model.

To allow for variance coefficients that are not constant and equal across assets, we can enrich the parametrization (\ref{eq:hepar}), for the variance part, with the logistic and linear parameterizations previously described.

By combining  the Log-BEKK for the variance coefficients and the HE-BEKK for the covariance coefficients, we obtain the Log-HE-BEKK model:
\begin{equation}
\begin{array}{rcl}
    \bm{A}_t&=&\bm{W}_t+a_1 G^*_t\\
    G^*_t&=&(\bm{I}_n+exp(-\phi(\bm{V}_{t-1}-c)))^{-1} 
    \end{array} \label{eq:loghepar}
\end{equation}
where $\bm{V}_{t-1}$ is a diagonal matrix containing the realized variances at time $t-1$. Both $\bm{W}_t$ and $G^*_t$ are PD, $a_1$ is nonnegative, which implies positive definiteness of $\bm{A}_t$ in (\ref{eq:loghepar}). Based on the previous results on the stability of the Log-HE model (Proposition \ref{prop3}), it is trivial to show that a sufficient condition for uniform contraction of the homogeneous component of the associated time-varying linear recursion is $(a+a_1+b)<1$.

By combining the Lin-BEKK for the variance coefficients and the HE-BEKK for the covariance coefficients, we obtain the Lin-HE-BEKK model:
(\ref{eq:tvbekk}):
\begin{equation}
    \bm{A}_t= \bm{W}_t-\alpha_a \bm{V}_{t-1} \\
    \label{eq:linhepar}
\end{equation}
 with $\alpha_a \ge0$. Again, $\alpha_a$ represents the attenuation effect described in subsection \ref{sec:linpar} (in this case referring only to the variances and not to the covariances). The parameterization (\ref{eq:linhepar}) can be considered a special case belonging to the family of models developed in \citet{Bauwens_Otranto:2025}.

\begin{proposition} The matrix $\bm{A}_t$ in (\ref{eq:linhepar}) is PSD if and only if $\alpha_a\le \min(\bm{\xi}^*)$, where $\bm{\xi}^*$ is the vector containing the minimum eigenvalues of the matrix $\bm{V}_{t-1}^{-1/2} \bm{W}_t \bm{V}_{t-1}^{-1/2}$ for each  $t$. A sufficient condition for the matrix $\bm{A}_t$ in (\ref{eq:linhepar}) to be PSD is $\alpha_a\le a \varphi\min(\bm{\xi}^{**})/exp(\varphi)$, where $\bm{\xi}^{**}$ is the vector containing the minimum eigenvalues of the matrix $\bm{V}_{t-1}^{-1/2} \bm{P}_{t-1} \bm{V}_{t-1}^{-1/2}$ for each  $t$. \label{prop7}
\end{proposition}

\begin{proof}
Both $\bm{W}_t$ and $\bm{V}_{t-1}$ are PD matrices. We rewrite $\bm{A}_t$ in (\ref{eq:hepar}) as:

\[
{\bm V}_{t-1}^{1/2}\left(\bm{V}_{t-1}^{-1/2} \bm{W}_t \bm{V}_{t-1}^{-1/2}-\alpha_a \bm{I}_n \right) {\bm V}_{t-1}^{1/2}
\]
$\bm{A}_t$ is PSD if and only if

\begin{equation}
\left(\bm{V}_{t-1}^{-1/2} \bm{W}_t \bm{V}_{t-1}^{-1/2}-\alpha_a \bm{I}_n \right)\succeq \bm{0}_n \label{eq:nclin}
\end{equation}
Consequently, a sufficient and necessary condition for PSDness of $\bm{A}_t$ is that $\alpha_a$ must be less than or equal to the minimum eigenvalue of $\bm{V}_{t-1}^{-1/2} \bm{W}_t \bm{V}_{t-1}^{-1/2}$ for every $t$, i.e. $\alpha_a \le \min(\bm{\xi}^*)$. Considering that $\bm{W}_t$ depends on the two unknown coefficients $a$ and $\varphi$, this condition is difficult to enforce in the estimation routine. We can rewrite:
\[
\bm{W}_t=\frac{a}{exp(\varphi)} \exp^\odot(\varphi \bm{P}_{t-1}) 
\]
Considering that:
\begin{equation}
\exp^\odot(\varphi \bm{P}_{t-1})=\bm{j}_n\bm{j}_n'+\varphi \bm{P}_{t-1}+\frac{\varphi^2}{2!}\bm{P}_{t-1}^{\odot2}+\frac{\varphi^3}{3!}\bm{P}_{t-1}^{\odot3}+\dots \label{eq:hpexp}
    \end{equation}
where ${\odot} h$ represents the $h$-th Hadamard (element-wise) power. By Schur's theorem, every element on the right of (\ref{eq:hpexp}) is  PSD and
\[
\exp^\odot(\varphi \bm{P}_{t-1}) \succ \varphi \bm{P}_{t-1}
\]
Consequently:
\[
\frac{a}{exp(\varphi)} \bm{V}_t^{-1/2}\exp^\odot(\varphi \bm{P}_{t-1})\bm{V}_{t-1}^{-1/2} \succ \frac{a \varphi}{exp(\varphi)}\bm{V}_{t-1}^{-1/2}\bm{P}_{t-1}\bm{V}_{t-1}^{-1/2}
\]
From (\ref{eq:nclin}), we can derive as a sufficient condition for the PSDness of $\bm{A}_t$:
\[
\frac{a \varphi}{exp(\varphi)}\bm{V}_{t-1}^{-1/2}\bm{P}_{t-1}\bm{V}_{t-1}^{-1/2} \succeq \alpha_a \bm{I}_n
\]
so that the constraint guaranteeing the condition is:
\[
\alpha_a\le a \varphi\min(\bm{\xi}^{**})/exp(\varphi)
\]
\end{proof}

\begin{proposition}
Given the conditions of Proposition \ref{prop7}, a sufficient condition for uniform contraction of the homogeneous component of the associated time-varying linear recursion of the process (\ref{eq:tvbekk_ct}) with parameterization (\ref{eq:linhepar}) is $(a+b)<1$. \label{prop8}
\end{proposition}

\begin{proof}
The diagonal elements of $\bm{A}_t$ are $a-\alpha_a \gamma_{rr}<a$, while the off-diagonal elements are less than $a$. Consequently every element of the diagonal matrix $\bm{D}_t$ in Proposition \ref{prop_si} is not greater than $(a+b)$, so uniform contraction is obtained if  $(a+b)<1$.
\end{proof}

\section{Estimator Properties}\label{sec:prop}
The estimation of model (\ref{eq:tvbekk_ct}) carried out using a two-step M-estimation procedure. In the first step, the matrix $\bar{\bm{\Gamma}}$ is estimated; in the second step, all remaining parameters are estimated conditional on the first-step estimate. Under standard regularity conditions, replacing nuisance quantities with consistent first-step estimators preserves the consistency of the second-step estimator \citep{NeweyMcFadden94,White_1994}. The matrix $\bar{\bm{\Gamma}}$ can be consistently estimated by the sample mean of the realized covariance matrices. In the second step, $\kappa$, $b$, and the reduced set of parameters defining $\bm{A}_t$ are estimated by quasi maximum likelihood (QML). 

\citet{Asai_So:2021} developed QML estimation for CAW-BEKK models and established the consistency and asymptotic normality of the resulting estimators under suitable regularity conditions. See also \citet{Asai_So:2023} for the covariance-targeting case. 

Let us decompose $\bm{\Gamma}_t$ as follows:
\begin{equation}
    \bm{\Gamma}_t=\bm{S}_t^{1/2} \bm{U}_t \bm{S}_t^{1/2}
\end{equation}
where $\bm{S}_t$ is parameterized as in (\ref{eq:tvbekk_ct}).  Under the following assumptions:
\begin{enumerate}
    \item $\bm{S}_t$ is $\mathcal{F}_t$-measurable;
    \item $E(\bm{\Gamma}_t|\mathcal{F}_{t-1})=\bm{S}_t$
    \item $\bm{U}_t$ is independent on $\mathcal{F}_{t-1}$ and $\bm{U}_t \sim W(\nu^{-1}\bm{I}_n, \nu)$
    \end{enumerate}
the second-step estimation  can be performed by maximizing the log-likelihood:
\begin{equation}
    \sum_{t=2}^T \left[\delta_t-\nu/2\left(ln|\bm{S}_t|+tr(\bm{S}_t^{-1} \bm{\Gamma}_t)\right)\right] \label{eq:log-lik}
\end{equation}
with
\[
\delta_t=\frac{n\nu}{2} ln (\frac{\nu}{2})-\frac{n(n-1)}{4}ln(\pi)+\frac{\nu-n-1}{2}ln|\bm{\Gamma}_t|-\sum_{i=1}^n ln \Gamma\left(\frac{\nu+1-i}{2}\right)
\]
The log-likelihood (\ref{eq:log-lik}) is obtained without specifying the conditional distribution of the covariance matrices and can be interpreted as QML estimator.
Furthermore, the estimates of the model parameters do not depend on $\nu$, so we can exclude $\nu$ from the estimation procedure. This will not affect the estimates of the other parameters, which are consistent.\footnote{Similar results were obtained by \citet{BauwensBraioneStorti2016}, using a DCC model \citep{EngleDCC02} instead of the BEKK.
} This result is also supported by the simulation results in \citet{Asai_So:2021}, who found that, setting different values for $\nu$,  the estimates of the other parameters do not change. 

The properties established for the CAW-BEKK model cannot be readily extended to the proposed models. The two-step procedure may entail some loss of efficiency, and the process need not be stationary. Moreover, the asymptotic properties of the estimators require further investigation because the model combines approximate covariance targeting, based on a first-step estimate, with nonlinear specifications of  $\bm{A}_t$. A formal analytical study of these properties is beyond the scope of this paper. Instead, we conduct Monte Carlo experiments to provide empirical evidence on the estimators' behavior in finite samples and as the sample size increases.

\subsection{Monte Carlo Evidence}\label{sec:mc}
We conducted a Monte Carlo experiment with a data generating process represented by (\ref{eq:tvbekk_ct}) with $n=30$, using the four alternative parameterizations for $\bm{A}_t$. Additionally, we generated sets of covariance matrices constructed, as a benchmark, from a simple scalar BEKK  model (sc-BEKK):
\[
{\bm S}_t=\kappa\bar{\bm{\Gamma}}+a  \bm{\Gamma}_{t-1}+b \bm{S}_{t-1}.
\]
and from the HE-BEKK (\ref{eq:hepar}) of \cite{Bauwens_Otranto:2023}.
The data generating parameters are the estimated values, reported in Table \ref{tab:est}, for the application described in Section \ref{sec:appl}. For each specification, we generated 500 replications for each sample size, $T=100, 500, 1000, 2000$. The covariance matrices $\bm{\Gamma}_t$ were generated from the Wishart distribution with parameters $\bm{S}_t$ and $\nu=90$;\footnote{Since the conditional variance of the Wishart distribution is inversely proportional to $\nu$, we set $\nu=3n$, which provides a reasonable compromise between low and high variability.} The initial value of $\bm{S}_0$ was extracted from the realized covariance series of the application in Section \ref{sec:appl}.

Table \ref{tab:mc} reports the bias, root mean squared error (RMSE), and the smallest sample size for which normality is not rejected (based on a Jarque-Bera test at 1\% significance).

\begin{table}
\centering
\footnotesize
\setlength{\tabcolsep}{3pt}
\renewcommand{\arraystretch}{1.15}

\caption{Monte Carlo experiment results: bias, RMSE, and normality for different sample sizes and 500 replications for each experiment. The last column indicates the smallest sample size for which normality is not rejected, using the Jarque-Bera statistic at 1\% significance.} 
\label{tab:mc}

\begin{tabular}{
    l
    >{\centering\arraybackslash}p{1.0cm}
    c c
    c c
    c c
    c c
    >{\centering\arraybackslash}p{1.25cm}
}
\toprule

\textbf{Parameter} &
\textbf{True} &
\multicolumn{2}{c}{$T=100$} &
\multicolumn{2}{c}{$T=500$} &
\multicolumn{2}{c}{$T=1000$} &
\multicolumn{2}{c}{$T=2000$} & 
 \\

\cmidrule(lr){3-4}
\cmidrule(lr){5-6}
\cmidrule(lr){7-8}
\cmidrule(lr){9-10}

&
&
\textbf{Bias} & \textbf{RMSE}
&
\textbf{Bias} & \textbf{RMSE}
&
\textbf{Bias} & \textbf{RMSE}
&
\textbf{Bias} & \textbf{RMSE}
&\textbf{Normality from $T$=} \\

\midrule

\multicolumn{11}{l}{\textbf{sc-BEKK}} \\

$\kappa$ & 0.089
& -0.002 & 0.003
& -0.001 & 0.001
& 0.000 & 0.001
& 0.000 & 0.000
& 500 \\

$a$ & 0.381
& 0.027 & 0.029
& 0.005 & 0.006
& 0.003 & 0.003
& 0.001 & 0.002
& 500 \\

$b$ & 0.554
& -0.024 & 0.026
& -0.004 & 0.005
& -0.002 & 0.003
& -0.001 & 0.002
& 100 \\

\midrule

\multicolumn{11}{l}{\textbf{Log-BEKK}} \\

$\kappa$ & 0.144
& 0.005 & 0.007
& 0.001 & 0.002
& 0.000 & 0.002
& 0.000 & 0.001
& 500 \\

$a_0$ & 0.312
& 0.019 & 0.021
& 0.004 & 0.005
& 0.002 & 0.003
& 0.001 & 0.002
& 500 \\

$a_1$ & 0.163
& 0.023 & 0.028
& 0.005 & 0.006
& 0.002 & 0.003
& 0.001 & 0.001
& 500 \\

$\phi$ & 10.318
& 0.663 & 10.928
& -0.056 & 1.414
& -0.002 & 0.797
& 0.001 & 0.007
& \\

$c$ & 0.780
& 0.011 & 0.048
& 0.001 & 0.016
& 0.000 & 0.011
& 0.000 & 0.006
& 500 \\

$b$ & 0.355
& -0.050 & 0.058
& -0.010 & 0.012
& -0.005 & 0.006
& -0.002 & 0.002
& 500 \\

\midrule

\multicolumn{11}{l}{\textbf{Lin-BEKK}} \\

$\kappa$ & 0.111
& 0.002 & 0.419
& 0.000 & 0.168
& 0.000 & 0.113
& 0.000 & 0.047
& 100 \\

$\alpha_v$ & 0.477
& 0.039 & 4.465
& 0.008 & 0.943
& 0.004 & 0.498
& 0.002 & 0.262
& 500 \\

$\alpha_c$ & 0.338
& 0.022 & 2.366
& 0.004 & 0.516
& 0.002 & 0.274
& 0.001 & 0.169
& 500 \\

$\alpha_a$ & 1.439
& 0.174 & 22.91
& 0.036 & 4.921
& 0.019 & 2.701
& 0.008 & 1.317
& 500 \\

$b$ & 0.381
& -0.044 & 5.131
& -0.009 & 1.064
& -0.004 & 0.545
& -0.002 & 0.253
& 500 \\

\midrule

\multicolumn{11}{l}{\textbf{HE-BEKK}} \\

$\kappa$ & 0.109
& 0.006 & 0.008
& 0.001 & 0.002
& 0.001 & 0.002
& 0.000 & 0.001
& 100 \\

$a$ & 0.492
& 0.048 & 0.053
& 0.010 & 0.012
& 0.005 & 0.006
& 0.002 & 0.003
& 500 \\

$\varphi$ & 0.763
& 0.024 & 0.041
& 0.005 & 0.011
& 0.003 & 0.007
& 0.001 & 0.004
& 500 \\

$b$ & 0.365
& -0.059 & 0.066
& -0.013 & 0.015
& -0.006 & 0.008
& -0.003 & 0.004
& 500 \\

\midrule

\multicolumn{11}{l}{\textbf{Log-HE-BEKK}} \\

$\kappa$ & 0.131
& 0.006 & 0.009
& 0.001 & 0.002
& 0.001 & 0.002
& 0.000 & 0.001
& 100 \\

$a$ & 0.419
& 0.028 & 0.036
& 0.007 & 0.009
& 0.004 & 0.002
& 0.002 & 0.003
& 500 \\

$\varphi$ & 0.491
& -0.002 & 0.080
& 0.003 & 0.018
& 0.002 & 0.010
& 0.001 & 0.006
&  \\

$a_1$ & 0.069
& 0.016 & 0.040
& 0.002 & 0.007
& 0.001 & 0.003
& 0.000 & 0.002
& 500\\

$\phi$ & 15.062
& 240.8 & 1271.1
& 0.570 & 9.472
& -0.006 & 0.007
& 0.008 & 0.065
& \\

$c$ & 1.994
& -0.023 & 0.213
& 0.003 & 0.034
& 0.003 & 0.021
& 0.001 & 0.015
& 500 \\

$b$ & 0.350
& -0.054 & 0.062
& -0.011 & 0.014
& -0.006 & 0.007
& -0.003 & 0.004
& 500 \\

\midrule
\multicolumn{11}{l}{\textbf{Lin-HE-BEKK}} \\

$\kappa$ & 0.109
& 0.006 & 0.008
& 0.001 & 0.002
& 0.001 & 0.001
& 0.000 & 0.001
& 100 \\

$a$ & 0.492
& 0.048 & 0.054
& 0.010 & 0.012
& 0.005 & 0.006
& 0.002 & 0.003
& 1000 \\

$\varphi$ & 0.764
& 0.025 & 0.040
& 0.005 & 0.011
& 0.002 & 0.007
& 0.001 & 0.004
& 500 \\

$\alpha_a$ & 0.128
& 0.013 & 0.015
& 0.003 & 0.003
& 0.001 & 0.002
& 0.001 & 0.001
& 2000\\

$b$ & 0.362
& 0.013 & 0.015
& -0.013 & 0.015
& -0.006 & 0.008
& -0.003 & 0.004
& 1000 \\

\bottomrule
\end{tabular}

\end{table}

The results are consistent with the consistency of the estimators across all models, with a marked reduction in bias as $T$ increases; in particular, the bias of the $\kappa$ estimator converges to zero very quickly for all models. For $T=2000$ the maximum bias is 0.008 ($\alpha_a$ in the Lin-BEKK model). For small samples ($T=100, 500$),\footnote{In the case of daily financial time series, the sample size is very often larger than 1000.} the bias is, in absolute terms, less than 0.06, excluding the coefficients multiplied by the covariance matrices constructed to give the time variability of $\bm{A}_t$: $\phi$ in Log-BEKK and Log-HE-BEKK, and $\alpha_a$ in Lin-BEKK.

The results are also encouraging with respect to efficiency; despite the two-stage estimation procedure, the parameter estimators of the new class of models show a similar reduction in RMSE compared to the benchmark scalar BEKK. The main exceptions are again $\phi$ in Log-BEKK and $\alpha_a$ in Lin-BEKK; the former maintains a high RMSE up to $T=1000$, then decreases dramatically for $T=2000$, while the latter shows the highest RMSE (equal to 1.317) for $T=2000$. The coefficient $\phi$ in Log-HE-BEKK starts from a very high RMSE (240.8) for $T=100$, but decreases very quickly, reaching the value of 0.065 for $T=2000$. The Log-HE-BEKK model is the most flexible specification, featuring distinct nonlinear parameterizations for the variance and covariance parameters; this greater flexibility may come at the cost of increased estimation variability in small samples.

\begin{figure}[t]
\includegraphics[width=15cm, height=8cm]{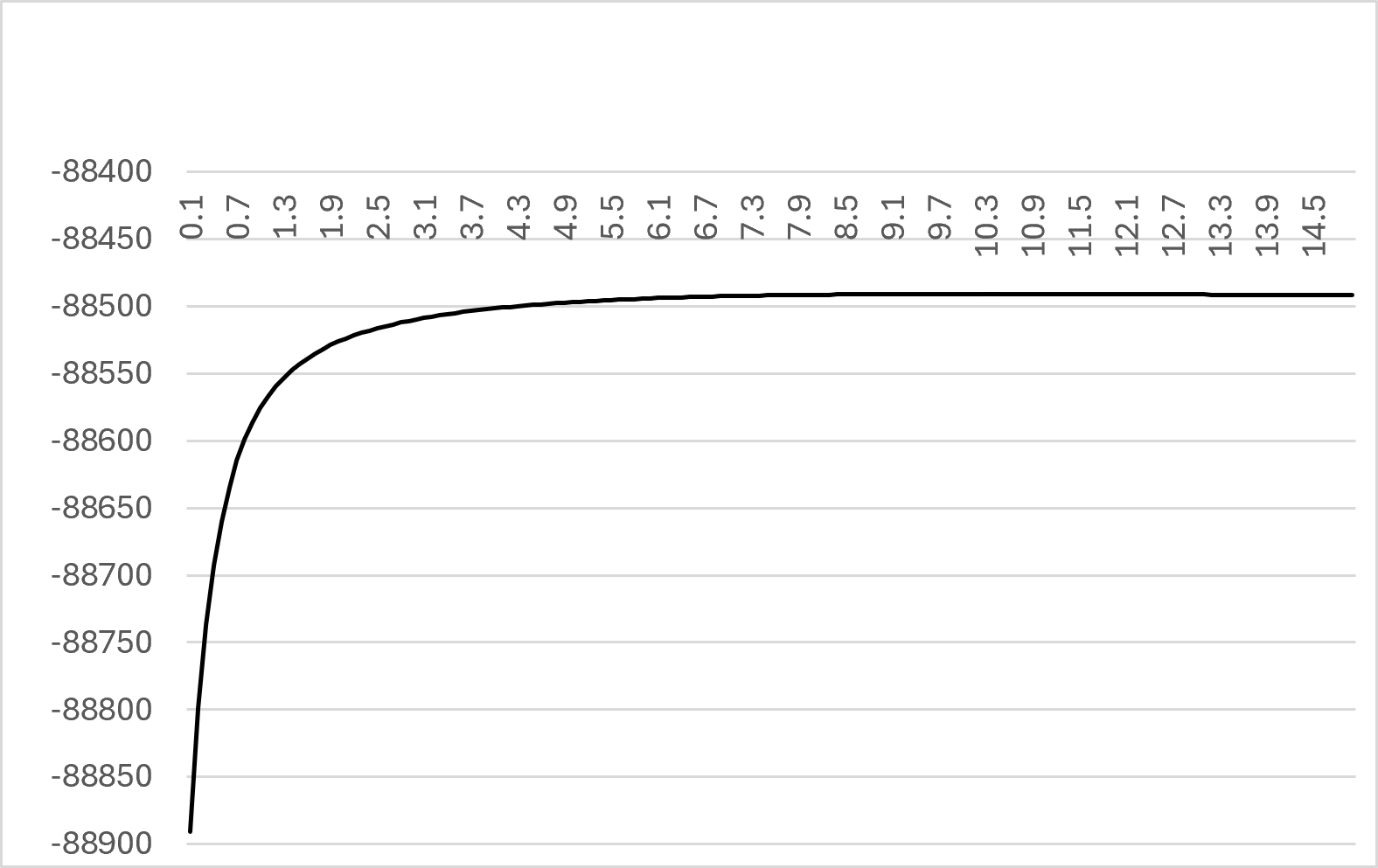}
\caption{DJIA dataset: Likelihood profile for the Log-BEKK model, at values of $\phi$.}\label{fig:lik_prof}
\end{figure}

In general, normality is verified even for small samples; most estimators achieve normality starting from $T=500$; this result is obtained at $T=100$ for the estimator of $\kappa$ in Lin-BEKK, Log-HE-BEKK and Lin-HE-BEKK, as for the HE-BEKK, while, in Lin-HE-BEKK, for   $a$ and $b$  starting from $T=1000$ and for $\alpha_a$ from $T=2000$.
The estimator of $\phi$ in Log-BEKK and Log-HE-BEKK does not achieve normality, nor does $\varphi$ in Log-HE-BEKK. The p-value of the Jarque-Bera statistic of $\varphi$ increases with $T$ and is unlikely to reject normality for values of $T$ greater than 2000, while the p-value of $\phi$ is always zero for both Log-BEKK and Log-HE-BEKK. A plausible explanation is the magnitude of the estimated parameter. As stated in subsection \ref{sec:st_par}, $\phi$ determines the shape of the logistic function, which does not change for large values of this parameter (saturation of the logistic matrix function). Figure \ref{fig:lik_prof} shows the likelihood profile with respect to $\phi$, computed on the real 30-variable dataset that will be analyzed in Section \ref{sec:appl} (DJIA dataset). It is clear that, for values of $\phi$ approximately greater than 7, the likelihood profile is flat, resulting in weak local identification.


\section{Empirical Application}\label{sec:appl}
We consider daily time series of annualized covariance matrices constructed from the 30 stocks included in the Dow Jones Industrial Average (DJIA), covering the period from January 2, 2015, to May 29, 2026. We use observations through the end of 2025 ($T=2765$) for in-sample estimation and model comparison, reserving the 102 observations from 2026 for out-of-sample evaluation.\footnote{As underlined in subsection \ref{sec:linpar}, constraints imposed to estimate the Lin-BEKK (Proposition \ref{prop5}) and Lin-HE-BEKK (Proposition \ref{prop7}) models are computed by considering the maximum eigenvalue of the realized covariance matrices and the minimum eigenvalue of the diagonal variance matrices, respectively, of the in-sample period. It is possible that these constraints could be different by adding the realized covariances of the out-of-sample period. In this application, this did not occur.} The latter is based on one-step-ahead predictions, with parameter estimates held fixed at their in-sample values. The data are freely available from the VOLARE platform  \citep[volare.unime.it;][]{VOLARE:2026}. 

We estimate six BEKK models: sc-BEKK, Log-BEKK, Lin-BEKK, HE-BEKK, Log-HE-BEKK, and Lin-HE-BEKK. The estimates are reported in Table \ref{tab:est}. We use QML estimation and calculate standard errors using the robust sandwich covariance estimator proposed by \citet{White82}.

\begin{table}[t]
\centering
\caption{Alternative BEKK model parameter estimates and corresponding log-likelihood, AIC and BIC, for the realized covariance series for the 30 assets that comprise the DJIA index. Robust standard errors in parentheses}\label{tab:est}
\begin{tabular}{lcccccc}
\hline
Parameters & sc & Log & Lin & HE & Log-HE&Lin-HE \\
\hline
$\kappa$ 
& 0.089 & 0.144 & 0.111 & 0.109 & 0.131&0.109 \\
& (0.018) & (0.028) & (0.021) & (0.019) & (0.026)&(0.020) \\

$a$
& 0.381 & 0.312 & & 0.492 & 0.419 &0.492\\
& (0.051) & (0.024) & & (0.068) & (0.040)&(0.069) \\

$b$
& 0.554 & 0.355 & 0.381 & 0.362 & 0.350&0.362 \\
& (0.055) & (0.098) & (0.094) & (0.086) & (0.085)&(0.087) \\

$a_1$
& & 0.163 & & & 0.069& \\
& & (0.049) & & & (0.026)& \\

$\phi$
& & 10.318 & & & 15.062& \\
& & (3.662) & & & (5.217)& \\

$c$
& & 0.780 & & & 1.994& \\
& & (0.078) & & & (0.150)& \\

$\varphi$
& & & & 0.763 & 0.491&0.764 \\
& & & & (0.117) & (0.044)&(0.116) \\

$\alpha_v$
& & & 0.477 & & & \\
& & & (0.074) & & & \\

$\alpha_c$
& & & 0.338 & & &\\
& & & (0.031) & & & \\

$\alpha_a$
& & & 1.439 & 2.880 & &0.128 \\
& & & (0.494) & (0.871) & &(0.083)\\

\hline
Log-Likelihood
& -89064.6 & -88490.9 & -88578.9 & -88527.1 & -88480.4&-88526.8 \\

AIC
& 64.45 & 64.04 & 64.10 & 64.06 & 64.03&64.06 \\

BIC
& 64.46 & 64.05 & 64.11 & 64.07 & 64.04&64.07 \\
\hline
\end{tabular}
\end{table}

In terms of log-likelihood, the proposed models outperform sc-BEKK. Log-HE-BEKK achieves the highest log-likelihood, which includes more parameters. However, the AIC and BIC results likewise indicate that Log-HE-BEKK performs well relative to the alternative specifications.

All parameter estimates are statistically significant at conventional levels, except for $\alpha_a$ in Lin-HE-BEKK, which is significant at the 10\% level. The estimated impact of new information—represented by the scalar $a$ in sc-BEKK and by the matrix $\bm{A}_t$ in the other models—differs substantially between the scalar and time-varying specifications. The persistence coefficient $b$ is notably larger in sc-BEKK, indicating a greater contribution from the lagged conditional covariance matrix $\bm{S}_{t-1}$ relative to the lagged realized covariance matrix $\bm{\Gamma}_{t-1}$, It is useful to compare the coefficient $a$ in Log-BEKK and Log-HE-BEKK. In the former, it represents the minimum value of the coefficient of lagged realized variances and covariances, augmented by the values of the multiple logistic function; in the latter, it represents the maximum value (in absolute terms) of the correlation coefficients, corrected (downward) by the HE function of lagged correlations, and the minimum value of the coefficient of realized variances, corrected (upward) by the values of lagged realized variance via the logistic function.

\begin{figure}[]
\includegraphics[width=15cm, height=18cm]{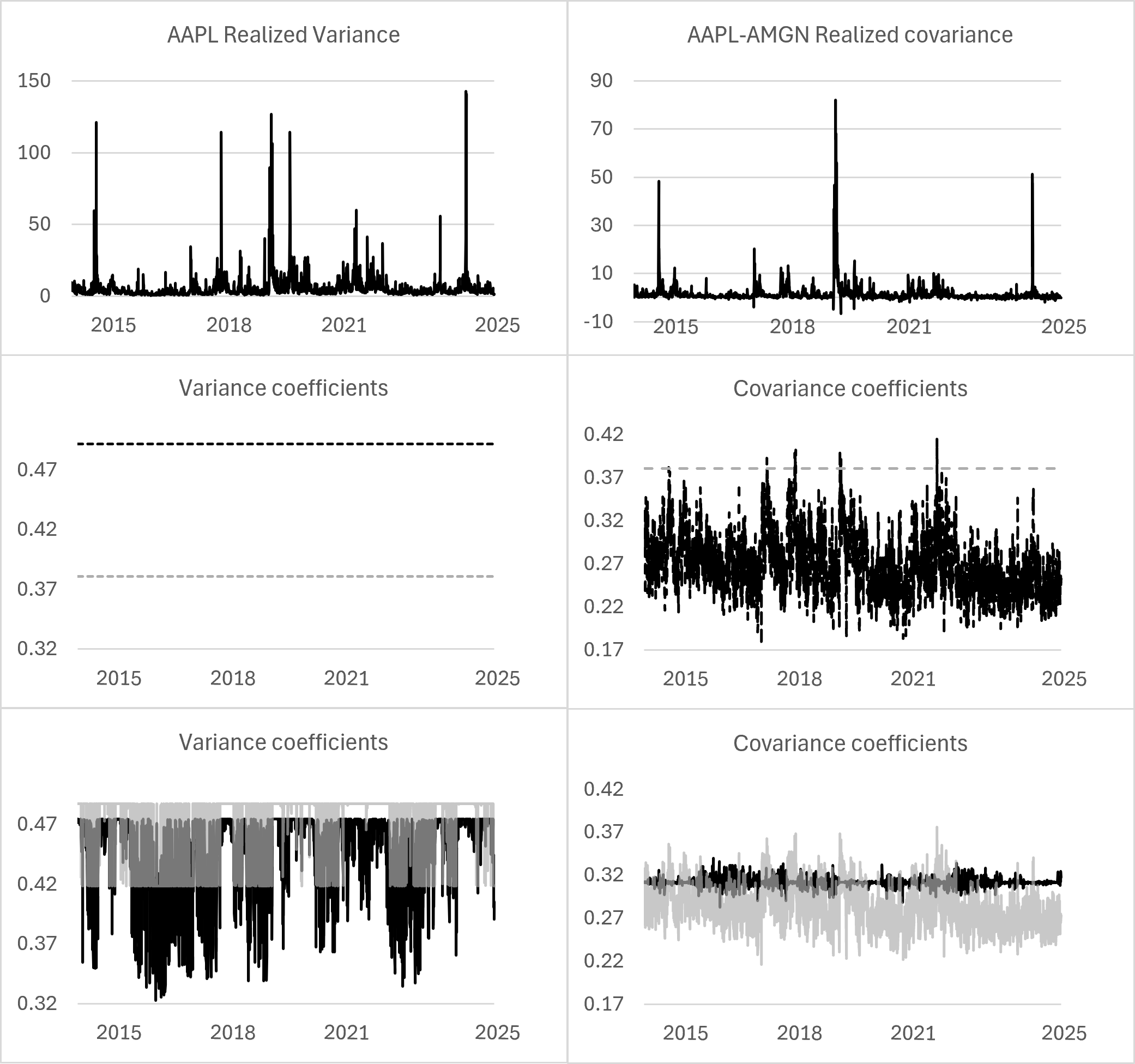}
\caption{DJIA dataset: Realized Variance of the AAPL asset and realized covariance of AAPL and AMGN (top panels) and corresponding time-varying coefficients  using sc-BEKK (gray dashed line) and HE-BEKK (black dashed line) (middle panel),   Log-BEKK (black continuous line) and Log-HE-BEKK (gray continuous line) (bottom panels), .}\label{fig:atv}
\end{figure}

The dynamics of the time-varying coefficients (for the lagged realized variance of the Apple Inc.-AAPL stock and the lagged covariance between AAPL and Amgen Inc.-AMGN), for the Log-BEKK, Log-HE-BEKK,  HE-BEKK and the constant coefficient $a$ of the sc-BEKK  are illustrated in Figure \ref{fig:atv}. HE-BEKK and sc-BEKK have constant variance coefficients, but the former is specific to variances, while the latter is common to variances and covariances. The news effect is smaller for covariances than for variances, so the specific coefficient of HE-BEKK is 0.11 points larger than that of sc-BEKK. Models with time-varying coefficients show similar movements, with $a$ representing the minimum value of the variance coefficients (0.312 for Log-BEKK and 0.419 for Log-HE-BEKK). For both models, the increase in coefficients depends on the logistic function, but only on the variances for Log-HE-BEKK and the variances and covariances for Log-BEKK; consequently, the highest values of the Log-HE coefficients are higher than those of Log (0.487 versus 0.475) and move within a narrow range.  Interestingly, the highest values of the Log-BEKK and Log-HE-BEKK variance coefficients correspond to the  highest volatility peaks (top left panel): the Chinese stock market crash of August 2015; the peak of November 20, 2018, following the resurgence of US-China trade tensions and growing concerns about global growth and monetary tightening; the effects of the 2020 COVID pandemic; and Trump's "Liberation Day" tariffs of April 2025.

In terms of covariance coefficients, the Log-HE and HE models exhibit very similar behavior, depending on the same HE structure. The Log-BEKK coefficients exhibit similar movements, but a small variability (ranging between 0.28 and 0.34), while the sc-BEKK model overestimates this coefficient.

The greater flexibility of the Log-HE-BEKK model explains its superior performance in terms of the likelihood function, AIC and BIC, as shown in Table \ref{tab:est}.
A further comparison is conducted using loss functions; we consider a statistical loss, the Quasi-Likelihood (QL),\footnote{We do not consider the multivariate version of the mean squared error (the Frobenius norm) as a loss function because it is very sensitive to sudden jumps, which are very common in this type of data.} defined as:
\[
QL=\frac{1}{T^*}\sum_{t=1}^{T^*} \left[ ln|\hat{{\bm S}}_t|+trace(\hat{{\bm S}}_t^{-1} \bm{\Gamma}_t)\right]
\]
and an economic loss function, the Global Minimum Variance \citep[GMV,][]{englecolacito2006}:
\[
GMV=\frac{1}{T^*}\sum_{t=1}^{T^*} \hat{{\bm w}}_t'\hat{{\bm S}}_t\hat{{\bm w}}_t\]
with weights  $\hat{{\bm w}}_t=\sqrt{n}\hat{{\bm S}}_t^{-1}\bm{j}_n/(\bm{j}_n'\hat{{\bm S}}_t^{-1}\bm{j}_n)$. $T^*$ represents the number of periods for which losses are calculated (referring to in-sample and out-of-sample periods).

Based on these statistics, for both the in-sample and out-of-sample spans, we  select the models with the best forecasting performance using the Model Confidence Set (MCS) approach of \citet{hansen2011mcs}.\footnote{For the sequential tests envisaged by the MCS approach, we use the semi-quadratic statistic and 10,000 bootstrap replications to calculate its standard deviation; see \citet{hansen2011mcs} for details. However, the alternative range statistic  provides similar results.} Of course, this exercise selects the models with the best fit in the in-sample analysis and the models with the best prediction performance for the out-of-sample case. \citet{hansen2011mcs} suggest a test significance level of 10\% or 25\% (the latter to make a more restricted selection).

The comparison is enriched by two alternative approaches, which are very popular in applications:  the vecHAR model \citep{ChiriacVoev_2011}, which can be estimated with the ordinary least squares method, and the exponentially weighted moving average (EWMA) model \citep{Riskmetrics96}, with a decay parameter equal to 0.96.

\begin{table}[htbp]
\centering
\caption{DJIA dataset: Loss functions for alternative models and MCS for in-sample and out-of-sample periods. The bold number indicates the corresponding model with the best performance;  numbers with $*$ indicate the corresponding models that perform non-significantly worse than the best model at the 10\% significance level. }
\label{tab:results}
{\begin{tabular}{lcccccccc}
\toprule
& \multicolumn{8}{c}{\textit{In sample}} \\
\cmidrule(lr){2-9}
& EWMA & vecHAR & sc & Log & Lin & HE & Log-HE&Lin-HE \\
\midrule
QL  & 65.977 & 67.667 & 64.446 &  64.031$^*$ & 64.095 &  64.057$^*$ & {\bf 64.023}&64.057$^*$ \\
GMV & 31.730 & 32.000 & 32.175 & 24.974 & 24.957 &  23.281 &  24.095&{\bf 23.278} \\
\midrule
& \multicolumn{8}{c}{\textit{Out of sample}} \\
\cmidrule(lr){2-9}
& EWMA & vecHAR & sc & Log & Lin & HE & Log-HE&Lin-HE \\
\midrule
QL  & 80.564  & 83.324 & 80.391 &  80.316 & 80.377 &  80.300 &  {\bf 80.287}&80.299$^*$ \\
GMV & 18.866 & 23.647 & 23.592 & 19.737 & 18.879 &  21.474 & 19.020&{\bf 18.010} \\
\bottomrule
\end{tabular}}
\end{table}

The two loss functions show similar results for both in-sample and out-of-sample periods. The Log-HE model achieves the best performance in terms of QL, but the Log, HE, and Lin-HE versions also fall into the best set at the 10\% significance level in the in-sample case, and Lin-HE in the out-of-sample case. It is worth noting that at the 25\% significance level, only Log-HE is in the  set of superior models, for both the in-sample and out-of-sample exercises.

Considering the GMV loss function, the Lin-HE version significantly outperforms all other models. Interestingly, in terms of AIC and BIC, Lin-HE does not improve on the HE-BEKK model (see Table \ref{tab:est}), but the changes to the variance coefficient it captures imply a significant improvement in GMV, especially in out-of-sample analysis.

 The new models show better performance than the classic EWMA, vecHAR, and sc-BEKK. This is an expected result for in-sample analysis, but is non-trivial for out-of-sample; the fact that \citet{HansenTimmerman2015} and \citet{XieYu2020} found that the simpler models performed better for out-of-sample analysis further emphasizes the performance of Log-HE-BEKK and Lin-HE-BEKK.
 
\section{Final Remarks}\label{sec_concl}
The availability of accurate measures of variances and covariances, such as realized covariance matrices, together with advances in computational power, has stimulated growing interest in covariance matrix models. When the objective is to model high-dimensional matrices, the curse of dimensionality compounds the challenge of ensuring that the model produces positive definite matrices.

We propose a family of parsimonious yet flexible models for realized covariance matrices. By estimating a relatively small number of parameters, the proposed models accommodate time-varying coefficients while ensuring the positive definiteness of the estimated conditional covariance matrices. The Monte Carlo results provide evidence on the finite-sample behavior of the estimators and their performance as the sample size increases, although some parameters exhibit non-normal sampling distributions when the likelihood is relatively flat. The empirical application indicates that the proposed models can outperform commonly used benchmark specifications.

Several extensions are possible. First, we develop the models within the BEKK framework, but the proposed approach can be extended to other baseline specifications, such as DCC \citep{EngleDCC02}, multivariate HAR \citep{Bauer_Vorkink_2011}, and DRD \citep{OhPatton2016}  models.

Second, we adopt a scalar specification for $\bm{B}_t$ in (\ref{eq:tvbekk}), motivated by previous empirical evidence. However, extending the model to a non-scalar $\bm{B}_t$ is straightforward. In this case, the sufficient condition for uniform contraction of the homogeneous component of the associated time-varying linear recursion in Proposition \ref{prop_si} becomes 
\[\sup_t \max_{i,j}|a_{ij,t}+b_{ij,t}|<1\] where $b_{ij,t}$ denotes the $(i,j)$-th element of $\bm{B}_t$.

Third, we use the standard exponential function in the Log-BEKK model and the Hadamard exponential function in the HE-type models. Either function could be used in both classes of models. We adopt the specifications that perform best in the empirical application.

Fourth, we use lagged realized covariance matrices-or lagged correlation matrices in the HE specifications-to drive the time variation in $\bm{A}_t$. More generally, this role could be assigned to any symmetric, time-varying positive semidefinite matrix $\bm{M}_{t}$. Furthermore, if $\bm{B}_t$ is parameterized in the same way as $\bm{A}_t$, the two coefficient matrices need not be driven by the same matrix $\bm{M}_{t}$. 


The proposed specifications can be viewed as multivariate covariance recursions with state-dependent coefficients. They are related to the literature on random-coefficient multivariate ARCH/GARCH models \citep[e.g.][]{FONGetal:2006} and, more generally, to stochastic recurrence equations with random coefficient matrices \citep{Bougerol_Picard:1992}.

\end{document}